\documentclass[aps,prd,a4paper,showpacs,floatfix,superscriptaddress,twocolumn]{revtex4-1}
\usepackage{graphicx}
\usepackage{color}
\usepackage{subfigure}
\usepackage{ulem}
\usepackage{epsfig}
\date{\today}
\begin{document}
TIFR-TH/11-29 .

\vspace{2mm}
\title{On the bulk viscosity of anisotropically expanding hot QCD plasma}
\author{Vinod Chandra}
\email{vinodc@theory.tifr.res.in}
\affiliation{Department of Theoretical Physics,
Tata Institute of Fundamental Research, Homi Bhabha Road, Mumbai-400005, India}
\begin{abstract}
The bulk viscosity, $\zeta$ and its ratio with the shear viscosity, $\zeta/\eta$ 
have been studied in an anisotropically expanding pure glue plasma in the presence of 
turbulent color fields. It has been shown that the 
anisotropy in the momentum distribution function of gluons,
which has been determined from a linearized transport equation
eventually leads to the bulk viscosity.  For the isotropic (equilibrium) state,
a recently proposed quasi-particle model of pure $SU(3)$ lattice QCD equation of state has been employed where 
the interactions are encoded in the effective fugacity. It has been argued that the interactions present 
in the equation of state, significantly contribute to the bulk viscosity. Its ratio with the shear viscosity is 
significant even at $1.5 T_c$. Thus, one needs to take in account the effects 
of the bulk viscosity while studying the hydrodynamic expansion of QGP in 
RHIC and LHC.

\pacs{25.75.-q; 05.20.Dd; 24.85.+p; 12.38.Mh}
\vspace{2mm}
\noindent {\bf Keywords}: Bulk viscosity; Shear viscosity; Quark-gluon plasma; Quasi-particle; Chromo-Weibel instability.
\end{abstract}
\maketitle

\section{ Introduction}
It is by now well established that Quark-gluon plasma (QGP) has been created in RHIC experiments, and   
is a strongly coupled fluid~\cite{expt}.There have been first few reports of QGP in Pb-Pb collisions $@ 2.76$ Tev in LHC\cite{lhc}, which reconfirm the formation of strongly coupled fluid. QGP at RHIC has shown a robust collective phenomenon,
 {\it viz.}, the elliptic flow\cite{flow_rhic}. In the heavy-ion collisions at LHC, there are other interesting flows, {\it viz.}, the dipolar, and the triangular flow which are sensitive to the initial collision geometry~\cite{flow_lhc}.
In this concern, we refer the reader to the very recent interesting studies~\cite{bhalerao,alice}, where these new kind of flows at LHC have been investigated.

The shear and bulk viscosities ($\eta$ and $\zeta$) characterize dissipative 
processes in the hydrodynamic evolution of a fluid. The former accounts for the entropy production 
due to the transformation of the shape of hydrodynamic system at a constant volume.
On the other hand, latter accounts for the entropy production at the constant rate 
of change of the volume of the system (in the context of RHIC the system stands for the fireball). 
These transport parameters serve as the inputs from the hydrodynamic evolution of the fluid. Their determination 
has to be done separately from a microscopic theory (either from a transport equation with 
appropriate force, collision and source terms or from the field theoretic approach using 
Green-Kubo formula). It has been found that QGP possess a very tiny value of the shear viscosity to entropy 
density ratio, $\eta/s$~\cite{shrvis}. On the other hand, bulk viscosity has achieved considerable attention in the context of QGP in RHIC after the interesting reports on its rising value close to the QCD transition temperature~\cite{khz1,khz2}. In the recent investigations, these transport coefficients are found to be sensitive to the interactions~\cite{chandra_eta1,chandra_eta2}, and nature of the phase transition in QCD~\cite{moore}.

The computation of transport coefficients in lattice QCD 
is a very non-trivial exercise, due to several uncertainties and 
inadequacy in their determination. Despite, there are a few first results computed from lattice QCD for bulk and 
shear viscosities~\cite{meyer,nakamura} which have observed a small value of $\eta/s$,  and a large value for $\zeta/s$ at RHIC.
While determining the behavior of the spectral function in~\cite{meyer}, a contribution coming from 
a $\delta$-function has not been taken in to account. This issue has been discussed extensively 
in~\cite{tmr}. The spectral density has been modified by incorporating the contributions from the 
$\delta$-function by Meyer in~\cite{meyer1}. However, a more refined lattice studies on $\eta$ and $\zeta$ 
are awaited in the near future with  less dependence on the lattice artifacts and uncertainties.
Subsequently, the possible impact 
of the large bulk viscosity of QGP in RHIC have been studied by several authors;
Song and Heinz~\cite{heinz} have studied, in detail, the interplay of 
shear and bulk viscosities in the context of collective flow in 
heavy ion collisions. Their study revealed that one can not simply 
ignore the bulk viscosity while modeling QGP in heavy ion collisions.
In this context, there are other interesting studies reported in the literature~\cite{den,raj1,hirano,raj,efaaf,pion,fries}. The role of bulk viscosity in freeze out phenomenon 
has been reported in~\cite{torri,hirano}. Effects of bulk viscosity in hadronic phase,  and in the hadron emission
have been reported in~\cite{boz}. There has been a wealth of recent literature on the computations of bulk
 viscosity in the context of cosmology~\cite{cosmo}, strange quark matter~\cite{sm}, and neutron stars~\cite{ns}.

The noteworthy point is that most of works devoted to study the hydrodynamic evolution of QGP,
employ constant value of $\eta/s$~\cite{shhydro} and $\zeta/s$~\cite{bulkhydro}.
This may not be desirable, in the light of experimental and phenomenological 
observation for QGP at RHIC. The work presented in this paper is an attempt to achieve, (i)
temperature dependence of transport coefficients, in particular, $\zeta$, (ii) 
to understand the large bulk viscosity of QGP.  In this study, 
we shall take inputs from the computations of bulk viscosity in quasi-particle models~\cite{sakai,quasi1}, 
and combine the understanding with a transport theory 
determination of $\zeta$ in the presence of Chromo-Weibel instabilities~\cite{bmuller,chromw}.
In this context the shear viscosity of QGP has already been addressed~\cite{bmuller,bmuller1,chandra_eta1,chandra_eta2}, and we find very interesting results. As it is well emphasized by Pratt~\cite{pratt} that there may be 
a variety of physical phenomena which can lead to viscous effects in QGP. Among them, in this paper,
we are particularly interested in the viscous effects which get contributions from the classical chromo-fields.

The idea adopted here is based on the mechanism, earlier proposed
to explain the small viscosity of a weakly coupled, but expanding hot QCD plasma~\cite{bmuller,bmuller1}. 
This mechanism is based on the particle transport theory in turbulent plasmas~\cite{dupree} which are 
characterized by strongly excited random field modes in the certain regimes of instability,
which coherently scatter the charged particles and thus reduce the rate of momentum
transport.This eventually leads to the suppression of the transport coefficients in plasmas.
This phenomenon in electro-magnetic (EM) plasmas has been studied in~\cite{niu}, and generalized by 
Asakawa, Bass and M\"{u}ller~\cite{bmuller} to the Non-Abelian plasma (QCD), and further employed 
for the realistic QGP EOS in~\cite{chandra_eta1,chandra_eta2}. As it is emphasized in~\cite{bmuller2}, 
the sufficient condition for the spontaneous
formation of turbulent, partially coherent fields is the presence of instabilities 
in the gauge fields due to the presence of charged particles.
This condition is met in both EM plasmas with an anisotropic momentum 
distribution~\cite{weibel} of charged particles and in QGP with an anisotropic distribution of thermal 
partons~\cite{sma}. Here, we shall argue that the similar mechanism can lead to a large bulk viscosity 
for the hot QCD plasma for the temperatures relevant at RHIC and heavy ion collisions at LHC.

The paper is organized as follows. In Sec. II, we present the general formalism to determine the transport parameters 
from a transport equation with a Vlasov term. We have neglected the collision and 
source term, while obtaining bulk viscosity. In Sec. III, we discuss the temperature 
dependence of bulk viscosity and its comparison with the shear viscosity. 
Finally, in Sec. IV, we present the conclusions and outlook.

\section{Transport parameters within a quasi-particle model}
The determination of transport coefficients requires modeling beyond the equilibrium properties,
in terms of the collision terms and other transport parameters, and also the nature
of perturbation to the equilibrium distribution. In particular, their determination within linearized
transport theory needs knowledge of EOS and the equilibrium momentum distribution functions of particles, 
which constitute the plasma. We shall first discuss the modeling of the EOS within a quasi-particle model.
The EOS chosen here is the pure $SU(3)$ gauge theory EOS~\cite{kar}. We subsequently discuss the setting 
up of the transport equation and the determination of $\zeta$. 
 
\subsection{The quasi-particle model}
Lattice QCD is the best, and most powerful technique to extract non-perturbative 
information on the equation of state for QGP~\cite{lat_eos,lat_eos1}.
Recently, we have proposed a quasi-particle model to 
describe the lattice data on pure $SU(3)$ gauge theory pressure (LEOS), and studied the 
bulk and transport properties of QGP~\cite{chandra_eta2}, which is utilized 
in obtaining the temperature dependence of bulk viscosity here. 
In this description, quasi-gluon distribution function extracted from LEOS possess the following  form,

\begin{equation} 
\label{eq1}
f_{eq}= \frac{z_g \exp(-\beta p)}{\bigg(1-z_g\exp(-\beta p)\bigg)}.
 \end{equation} 

It has further been argued\cite{chandra_eta2}
that the model is in the spirit of Landau theory of 
Fermi liquids. The connection with the Landau's theory is apparent from 
the single quasi-gluon energy, which gets non-trivial contributions from 
the quasi-particle excitations. The dispersion relation (single particle energy)
came out to be, 

\begin{equation}
\label{eq2}
 E_p=p+T^2\partial_T \ln(z_g).
\end{equation}

The main feature of the description is the 
mapping of strongly interacting LEOS in to a system of 
non-interacting/weakly interacting quasi-gluons (free up to the 
temperature dependent fugacity, $z_g$ which encodes all the interactions, and the dispersion relation
in Eq.(\ref{eq2})). This enables us to tackle highly non-trivial strong interaction 
in QCD in a very simplified manner while studying the properties of 
QGP. Interestingly, Eq.(\ref{eq2}), which  is obtained from the thermodynamic 
definition of the energy-density in terms of Grand-canonical QCD partition functions,
ensures the thermodynamic consistency in hot QCD, and reproduces the lattice results on the trace anomaly correctly.
This is also true  for the recently proposed quasi-particle model which describes the $(2+1)$-flavor 
lattice QCD~\cite{vinod}.

This quasi-particle understanding of hot QCD has been 
quite successful in describing the realistic QGP equations of state, and
in investigating the bulk and transport properties of QGP~\cite{vinod_quasi,vinod_quasi1,chandra_eta1,chandra_eta2}.
We shall utilize Eqs.(\ref{eq1}) and (\ref{eq2}) to determine the bulk and shear viscosities
within transport theory framework here. Note that there are other quasi-particle approaches to describe
lattice QCD EOS based on effective thermal masses for quasi-partons~\cite{kamp,pesh,bannur,rebhan,thaler},
approaches based on Polyakov loop~\cite{polyakov}, and quasi-particle models with gluon 
condensate~\cite{ella,casto}. Recently, transport coefficients for QGP within the effective mass models 
in the relaxation time approximation have been reported in~\cite{bluhm,aks}. As argued in~\cite{vinod_quasi1}, 
our model is distinct from all these approaches, but equally successful  in describing the 
thermodynamics of QGP.

\subsection{Determination of the transport coefficients}
We now consider the important physical quantities, 
the bulk viscosity, $\zeta$, its ratio with entropy density, 
$\zeta/s$. For the entropy density, we again utilize the 
lattice results quoted in~\cite{chandra_eta2}. These quantities are very crucial to understand the QGP
in RHIC. Their determination requires knowledge of the collisional properties
of the medium when it is perturbed away from equilibrium.
To determine these quantities, we adopt approach of 
~\cite{bmuller,bmuller1,chandra_eta2}. The shear viscosity had been determined 
in~\cite{chandra_eta2}, which we shall utilize to study the ratio $\zeta/\eta$
in the later part. Here, we consider $\zeta$ and determine it 
from a transport equation. 

The determination of bulk viscosity has been done in a  multi-fold way.
Firstly, we need an appropriate modeling of distribution function for the equilibrium state.
Secondly, we need to set up an appropriate transport equation to determine the 
form of the perturbation to the distribution function. These two steps eventually determine the 
bulk viscosity. For the former step, we employ the quasi-particle model for 
LEOS discussed earlier. We shall leave 
the analysis in the case of full QCD for future investigations.

The bulk viscosity has two contributions same as the shear viscosity in~\cite{bmuller}, (i)
from the Vlasov term which captures the long range component
of the interactions, and (ii) the collision term, which
models the short range component of the interaction. 
Here, we shall only concentrate on the former case. 
The determination of shear and bulk viscosities 
from an appropriate collision term will be a matter 
of future investigations. Importantly, the analysis adopted here is 
based on weak coupling limit in QCD, therefore, the results are shown beyond 
$1.3 T_c$ assuming the validity of weak coupling results for QGP there.

\subsubsection{Formalism}
Let us first briefly outline the standard procedure of determining transport coefficients in 
transport theory~\cite{landau,bmuller}. The bulk and shear viscosities, $\zeta$ and $\eta$ of QGP in terms of 
equilibrium parton distribution functions are obtained by comparing the 
microscopic definition of the stress tensor with the macroscopic definition of the viscous stress tensor. 
The microscopic definition of the stress tensor is,

\begin{equation}
\label{eq8}
 T_{i k}=\int \frac{d^3p}{(2\pi)^3 E_p} p_i p_k f(\vec{p},\vec{r}).
\end{equation}

On the other hand, macroscopic expression for the viscous stress tensor is given by,

\begin{equation}
\label{eq9}
T_{i k}=P\delta_{i k}+\epsilon u_i u_k-2\eta(\nabla u)_{i k} -\zeta \delta_{i k} \nabla\cdot\vec{u},
\end{equation}

where $(\nabla u)_{i k}$ is the traceless, symmetrized
velocity gradient, and $\nabla\cdot\vec{u}$ is the divergence of the fluid velocity field.
$E_p$ accounts for the dispersion relation.
To determine $\zeta$ an $\eta$, one writes the gluon distribution function as

\begin{equation}
\label{eqn9c}
f(\vec{p},\vec{r})=\frac{1}{{z_{g}}^{-1}\exp(\beta u\cdot p-f_1(\vec{p},\vec{r}))- 1}.
\end{equation}

Assuming that  $f_1(\vec{p},\vec{r})$ is a small perturbation to the equilibrium distribution, we expand $f(\vec{p},\vec{r})$ and keep the linear order term in $f_1$;
this leads to,

\begin{eqnarray}
 f(\vec{p},\vec{r})&=&f_0({\bf p})+\delta f(\vec{p},\vec{r})\nonumber\\
&=&f_0({\bf p})\bigg(1+f_1(\vec {p},\vec{r})(1+ f_0({\bf p}))\bigg),
\end{eqnarray}
 where $f_0({\bf p})$ is the isotropic distribution function, as we shall see that this will be 
same as the equilibrium thermal distribution function of the quasi-gluons, in the 
rest frame of the fluid.
As discussed in \cite{chandra_eta2}, $\zeta$ and $\eta$ are determined by 
taking the following form of the perturbation $f_1$,

\begin{equation}
\label{eq11}
 f_1(\vec{p},\vec{r})=-\frac{1}{E_p T^2} p_i p_j \bigg({\Delta_1}(p)\nabla u)_{i j}+{\Delta_2}(\vec{p})(\nabla.u)\delta_{ij}\bigg),
\end{equation}

 where  the dimensionless functions $\Delta_{1}(p), \Delta_{2}(\vec{p})$  measure the deviation from the 
equilibrium configuration. $\Delta_1(p)$, $\Delta_2(\vec{p})$,  lead to $\eta$ and $\zeta$ respectively. 
Note that  $\Delta_1(p)$ is an isotropic function of the momentum in contrast to $\Delta_2(\vec{p})$, 
which is an anisotropic in momentum $\vec{p}$.

Since $\zeta$ and $\eta$ are  Lorentz scalars; they may be evaluated conveniently in the local rest frame. In the local rest frame of the fluid $f_0\equiv f_{eq}$.
Considering the a boost invariant longitudinal flow, 
$\nabla\cdot u=\frac{1}{\tau}$ and, $(\nabla u)_{i j} = \frac{1}{3\tau} diag (-1, -1,2)$,  in the local rest frame,
we find that  $f_1(p)$ takes the  form,

\begin{equation}
\label{eq9a}
f_1(\vec{p})=-\frac{{\Delta_1}(p)}{E_p T^2\tau}\bigg(p_z^2-\frac{p^2}{3}\bigg)-\frac{{\Delta_2}(\vec{p})}{E_p T^2\tau}
p^2,
\end{equation}

where $\tau$ is the proper time($\tau=\sqrt{t^2-z^2}$). 
The shear and bulk viscosities  
are obtained in terms of entirely unknown function $\Delta_1(p)$ and $\Delta_2(\vec{p})$ as,

\begin{equation}
\label{eq10}
\eta=\frac{\nu_g}{15 T^2}\int \frac{d^3 p}{8\pi^3} \frac{p^4}{E_p^2} \Delta_1(p)f_{eq}(1+f_{eq}),
\end{equation}

\begin{equation}
\label{xi}
\zeta=\frac{\nu_g}{3 T^2} \int \frac{d^3 p}{8\pi^3} \frac{p^2}{E_p^2} (p^2-3 c^2_s E_p^2)\Delta_2(\vec{p}) f_{eq}(1+f_{eq}).
\end{equation}

In these expressions, $\nu_g\equiv 2(N_c^2-1)$ is the degrees of freedom.
Notice that while obtaining the expression for the bulk viscosity, we have 
exploited the Landau-Lifshitz condition for the stress energy tensor. 
The factor $-(3 c^2_s E_p^2)$ in the {\it rhs} of Eq.(\ref{xi}) is coming only because of that. For details, we refer the reader to \cite{quasi1}.
 
The determinations of $\Delta_1(p)$, and $\eta$ have already been done in \cite{chandra_eta1,chandra_eta2}. 
We shall utilize these results to fix the temperature dependence of $\zeta$ in the later part of the analysis. 
Now, we shall focus on the determination of the unknown function $\Delta_2(\vec{p})$ and $\zeta$.

\subsubsection{Determination of $\Delta_2(\vec{p})$}
For simplicity, we consider the purely chromo-magnetic plasma for our analysis.
The modeling of transport equation for full chromo-electromagnetic plasma is
straight forward\cite{bmuller,chandra_eta1,chandra_eta2} and differs by simple factors.
Here, we only quote the mathematical form of
the drift term and the Vlasov term (For details see \cite{bmuller,chandra_eta1}).

The drift term in the transport equation for the full chromo-electromagnetic plasma for LEOS is obtained as,

\begin{eqnarray}
\label{eqd} 
(v\cdot\partial)f_{eq}(p)&=&f_{eq}(1+f_{eq})\bigg[\frac{p_i p_j}{E_p T} (\nabla u)_{ij}\nonumber\\&-&\frac{m^2_D <E^2> \tau^{el}_m E_p}{3T^2 {\partial {\mathcal E}}/{\partial T}}+(\frac{p^2}{3 E^2_p}-c^2_s)\frac{E_p}{T}(\nabla\cdot\vec{u})\bigg],\nonumber\\\end{eqnarray} 
where $c^2_s$ is the speed of sound. The other notations are kept same as in~\cite{chandra_eta2}.
Note that $<E^2>$ stands for the chromo-electric field, $\tau_{el}$ relaxation time associated with the instability\cite{bmuller}. In Eq.(\ref{eqd}), first term contributes to the shear viscosity, second term contributes to the thermal conductivity, and the third term contributes to the bulk viscosity. Since, we are considering the purely chromo-magnetic plasma, so the 
second term will not be present.   

On the other hand the force term (Vlasov term) which we denote as ${\bf V}_A$,  is obtained as\cite{bmuller,chandra_eta2} follows, 

\begin{equation}
{\bf V}_A=\frac{g^2 C_2}{2(N_c^2-1) E_p^2}<B^2> \tau_m {\bf L^2},
\end{equation}  

where $C_2=N_c$, $<B^2>$ denotes chromo-magnetic field, $\tau_m$ is the times scale associated with 
instability in the field,  and the operator ${\bf L^{2}}$ is
 
\begin{eqnarray}
{\bf L^2}&=&-(\vec{p}\times\partial_{\vec p})^2+(\vec{p}\times\partial_{\vec p})\vert_z^2\nonumber\\
      &&\equiv -(L^{p})^2+({L^{p}}_{z})^2.\nonumber\\
\end{eqnarray}

Since ${\bf L^2}$ contains angular momentum operator $L^{p}$, therefore it gives non-vanishing 
contribution while operating on an anisotropic function of $\vec{p}$. It will always lead to the vanishing 
contribution while operating on an isotropic function of $\vec{p}$. Therefore,  ${\bf V}_A f_{eq}\equiv 0$. 
Now, we write the transport equation containing only those terms which contribute to 
bulk viscosity $\zeta$ as,

\begin{eqnarray}
(\frac{p^2}{3 E^2_p}&-&c^2_s)\frac{E_p}{T}(\nabla\cdot\vec{u})f_{eq}(1+f_{eq})
\nonumber\\=&&\frac{g^2 C_2}{3(N_c^2-1) E_p^2}<B^2> \tau_m {\bf L}^2 \ f_1(\vec{p},\vec{r})f_{eq}(1+f_{eq}).\nonumber\\
\end{eqnarray}

Substituting for $f_1$ in term of the unknown function $\Delta_2(\vec{p})$ and rearranging 
above equation, we obtain a differential equation for $\Delta_2(\vec{p})$ as,

\begin{equation}
{\bf L^2} \Delta_2(\vec{p})=\frac{2(N_c^2-1)T E_p^2}{N_c g^2 <B^2> \tau_m p^2 }(\frac{p^2}{3}-c^2_s\ E_p^2)\nonumber\\
\end{equation}

Now, using the fact that ${\bf L^2}$ only operates on the anisotropic function 
of $\vec{p}$, we can write,

\begin{equation}
\label{delp}
\Delta_2(\vec{p})=\frac{2(N_c^2-1)T E_p^2}{N_c g^2 <B^2> \tau_m p^2 }(\frac{p^2}{3}-c^2_s\ E_p^2)\times g(\vec{p}),
\end{equation}

where $g(\vec{p})$ can be determined from the following condition,

\begin{equation}
{\bf L^2}\ g(\vec{p})= 1,
\end{equation}

which leads to,

\begin{equation}
\label{gp}
 g(\vec{p})=\frac{1}{2}\ln(\frac{p_x^2+p_y^2}{p^2_0})\equiv \ln(\frac{p_T}{p_0}).
\end{equation}

Since, at high temperature average value of the 
energy is $3\ T$. Employing equipartition theorem for relativistic massless 
gas, we obtain $p^2_0=6 \ T^2$. Substituting Eq.(\ref{gp}) in Eq.(\ref{delp}), we obtain,

\begin{eqnarray}
\label{1}
\Delta_2(\vec{p})=\frac{2(N_c^2-1)T E_p^2}{N_c g^2 <B^2> \tau_m p^2 }(\frac{p^2}{3}-c^2_s\ E_p^2)\ \ln(\frac{p_T}{p_0}) 
\end{eqnarray}

The determination of bulk viscosity is incomplete unless we know not only the 
temperature dependence of the speed of sound square, $c_s^2$, and  the 
the collective contributions of quasi-particle to the single particle energy, $T^2\partial_{T}\ln(z_g)$ but also the quantity $g^2 <B^2> \tau_m$.  

We determine first two quantities using the quasi-particle model.
As from Ref.\cite{chandra_eta2}, the trace anomaly in terms of effective quasi-particle number density and
effective gluon fugacity reads,

\begin{equation}
\frac{(\epsilon-3P)}{T^4}=\frac{{\cal N}_g}{T^3} \lbrace T\partial_T \ln(z_g)\rbrace.
\end{equation}

The thermodynamic quantities can be obtained using the well known thermodynamic relations.
In particular, the energy density and the entropy density was shown to be in almost 
perfect agreement with the lattice data~\cite{chandra_eta2}. We determine, $c^2_s$ by employing 
a method reported in~\cite{gupta}. The temperature dependence is shown in 
Fig. 1.

To relate the denominator of Eq. (\ref{1}) to the gluon quenching parameter, $\hat{q}$ we go the light cone frame.
In this frame, Eq.(\ref{1}) can be rewritten as,

\begin{equation}
\label{eqz}
\Delta_2(\vec{p})=\frac{4(N_c^2-1)T E_p^2}{N_c g^2 <E^2+B^2> \tau_m p^2 }(\frac{p^2}{3}-c^2_s\ E_p^2)\ \ln(\frac{p_T}{p_0}). 
\end{equation}

The gluon quenching parameter, $\hat{q}$ is related with the denominator of {\it rhs} of the above equation as~\cite{bmuller1},
\begin{equation}
\hat{q}=\frac{2 g^2 N_c}{3(N_c^2-1)} <E^2+B^2> \tau_m .
\end{equation}

Now,employing Eq.(\ref{eqz}) in Eq.(8), we obtain the 
$\zeta$ as,

\begin{eqnarray}
\label{eqxi}
\zeta &=& \frac{(N_c^2-1)}{3 T \pi^2 \hat{q}} \int_0^\infty \int_{-\infty}^{\infty}\  p_{T} dp_{T} dp_{z} (\frac{p^2}{3}-c^2_s\ E_p^2)^2\times\nonumber\\ && \ln(\frac{p_T}{p_0})\times f_{eq}(1+f_{eq}).
\end{eqnarray}

On the other hand, if we employ the results of \cite{chandra_eta2} for $\Delta_1(p)$ in Eq.(10) for $\eta$, 
we obtain,

\begin{equation}
\label{eqet}
\eta= \frac{T^6}{\hat{q}} \frac{64(N_c^2-1)}{ 3\pi^2} PolyLog[6,z_g],
\end{equation}
where $N_c=3$ and $PolyLog[6,z_g]=\sum_{k=1}^\infty \frac{{z_g}^k}{k^6}$.
\begin{figure}
\vspace{2mm}
\includegraphics[scale=.40]{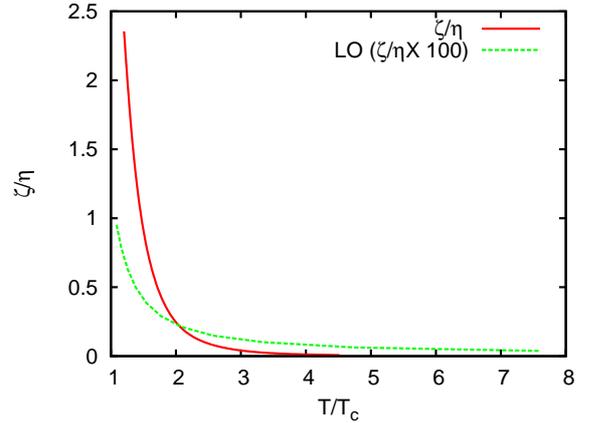} 
\caption{\label{bulks} (Color online) The ratio of bulk viscosity, $\zeta$ to the shear viscosity, $\eta$ 
as a function of temperature.  The leading order (LO) result of $\zeta/\eta$ has been obtained from the data taken from Refs.\cite{chen_bulk,chen_shear}, and shown as dashed line. For the sake of comparison, we have multiplied the leading order
$\zeta/\eta$ by a factor of hundred.}
\vspace{2mm}
\end{figure}

 Now scaling, all the quantities in the integrand in Eq.(\ref{eqxi}) by $T$, and rewriting Eq.(\ref{eqet}) in
 the form given below, we obtain,

\begin{equation} 
\label{eqhat}
 \zeta=\frac{T^6}{\hat{q}} I_{1}(T/T_c);\ \eta=\frac{T^6}{\hat{q}} I_{2}(T/T_c),
\end{equation}
where $I_{1}(T/T_c)$, is evaluated by integrating the {\it rhs} of Eqs.(\ref{eqxi}) numerically, 
and $I_{2}(T/T_c)\equiv\frac{8^3}{3\pi^3} PolyLog[6,z_g]$. The $T/T_c$ scaling of these quantities is
 coming from the temperature dependence of the effective gluon fugacity, $z_g$. Here, $T_c$ is taken to
 be  $0.27\ GeV$~\cite{zantow}. Clearly, the quantity which can be determined unambiguously in our approach is the 
ratio $\zeta/\eta\equiv {I_{1}(T/T_c)}/{I_{2}(T/T_c)}$. 

In the recent past, Chen {\it et. al}~\cite{chen_bulk,chen_shear} have 
computed the leading order shear and bulk viscosities for  purely gluonic plasma. This is nothing but the collisional contribution
to these transport parameters for a gluonic plasma. It is to be instructive to compare the results on $\zeta/\eta$ obtained in the present work with those reported in~\cite{chen_bulk}. This has been shown in Fig. 1, where both the 
results on $\zeta/\eta$ are plotted as a function of $T/T_c$. Note that while obtaining the temperature dependence of the 
ratio $\zeta/\eta$, we have employed the two-loop expression for the 
running coupling constant at finite temperature quoted in \cite{chen_bulk}. Quantitatively the ratio $\zeta/\eta$ is much smaller than what we have obtained from the diffusive Vlasov term. If we compare the two curves on the ratio $\zeta/\eta$ shown in Fig. 1, we find that in contrast to our prediction on $\zeta/\eta$, the leading order result suggests the near conformal picture of hot QCD even at lower temperatures.

Next, we discuss the interplay of the two contributions to the bulk viscosity, {\it viz.}, 
the anomalous, and the leading order (collisional). As it is emphasized in~\cite{bmuller}, these two contributions 
for $\eta$ are inverse additive. Their inverse additivity has been argued from the additivity of various rates 
in the hot QCD medium. In the case of weak coupling, the former is predominant. It seems that a similar 
additivity of the inverse of two contributions to $\zeta$,{\it viz.} (denoted as $\zeta_a$ and $\zeta_c$ respectively) may perhaps be valid. This could be understood as follows: since $\zeta_a$ is inversely proportional to the $\hat{q}$ (transport rate), on the other hand collisional $\zeta_c$ will be inversely proportional to the collision rate. Following the argument previously mentioned, one may write, $\zeta_T^{-1}=\zeta_{a}^{-1}+\zeta_c^{-1}$, where $\zeta_T$ denotes the total bulk viscosity. This inverse additivity of $\zeta_a$, and $\zeta_c$
at weak coupling, suggests that the collisional bulk viscosity (leading order) will dominate over the anomalous one, since the former is quantitatively much smaller than the latter. However, one has be very cautious while comparing these two contributions for the temperature ranges relevant for QGP at RHIC. This is because of the strongly coupled fluid like picture of QGP. At this moment, we do not know whether the inverse additivity of $\zeta$ will be followed at the temperatures which are closer to $T_c$ or not. This is a very crucial issue, and will require much deeper investigations, which is beyond the scope of the present work. Henceforth, we denote the anomalous bulk viscosity as $\zeta$ dropping the subscript, {\it a}. 

We now proceed to discuss the temperature
dependence of $\zeta/\eta$ and $\zeta/s$.  

\begin{figure}
\vspace{2mm}
\includegraphics[scale=.40]{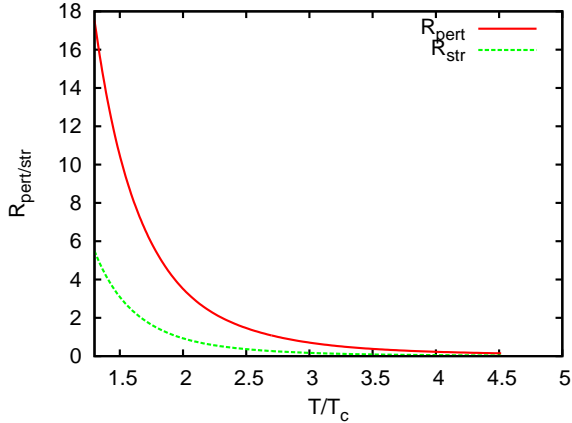} 
\caption{\label{bulkpert} (Color online) 
Comparison of the ratio $\zeta/\eta$ with the perturbative QCD, and strongly coupled theories. The quantities 
$R_{pert}$, and $R_{str}$ are defined in terms of the ratios $\frac{\zeta}{\eta (c_s^2-\frac{1}{3})^2}$, and $\frac{\zeta}{\eta (c_s^2-\frac{1}{3})}$ respectively. Here, {\it pert} stands for perturbative QCD, and {\it str} stands for the strongly coupled near conformal theories.}
\vspace{2mm}
\end{figure}

\begin{figure}
\vspace{2mm}
\includegraphics[scale=.40]{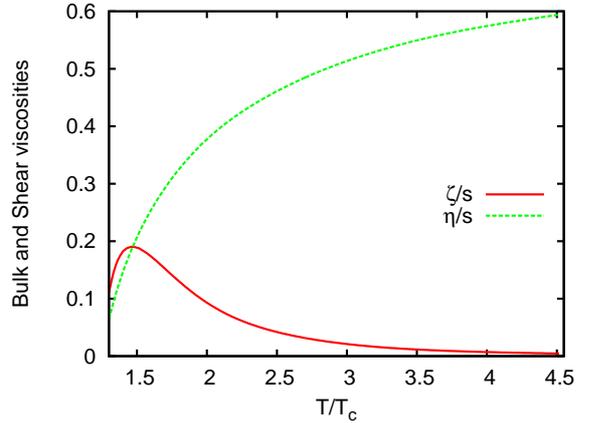} 
\caption{\label{bulk} (Color online) $\zeta/s$ and $\eta/s$ as a function of temperature. The dashed (green) line denotes $\zeta/s$ and 
solid(red) line denotes the $\eta/s$.}
\vspace{2mm}
\end{figure}

\subsection{Temperature dependence of  $\zeta/\eta$ and $\zeta/s$}
In our analysis, determination of the temperature dependence of  $\zeta$ and $\eta$ is 
incomplete, without the knowledge of the temperature dependence of $\hat{q}$ in QGP.  This issue was addressed by fixing the temperature dependence of $\hat{q}$ by calculating the soft part of the energy density and the 
relaxation time associated with the instability of chromo-fields~\cite{chandra_eta2}.
To do that we take inputs from the phenomenological values of $\hat{q}$,
which is known at a particular temperature~\cite{hatq}. Here, we have utilized the same transport equation
and quasi-particle model developed for pure $SU(3)$ lattice QCD EOS, as in~\cite{chandra_eta2}.
Therefore, we employ the temperature dependence of $\eta/s$ to obtain the temperature 
dependence of $\zeta/s$. This is quite easier to do, since the ratio, $\zeta/\eta$ can easily be 
obtained from Eq. (\ref{eqhat}).  

The temperature dependence of $\zeta/\eta$ is shown in Fig. 1, $\zeta/\eta$ relative to perturbative
 QCD prediction~\cite{arnold}, and strongly coupled near conformal gauge theories~\cite{confo} is shown 
in Fig. 2. On the other hand, $\zeta/s$ and $\eta/s$ are shown together in Fig. 3. Let us discuss their behavior
 one by one. From Fig. 1, it is clear that  $\zeta/\eta$ is equally significant while studying the hydrodynamic 
evolution of hot QCD matter until we reach $T= 2 T_c$. As we go to the higher temperatures the ratio further
 decreases and eventually vanishes when $c_s^2=\frac{1}{3}$ and the dispersion relation $E_p=p$. Quantitatively, $\zeta/\eta \sim 2.3$ at $1.3 T_c$; $1.0$ at $1.5 T_c$; $0.2$ at $2.0 T_c$.
Therefore, for  $T\geq 2.5 T_c$, one can ignore $\zeta$ over $\eta$. In other words, the hot QCD becomes almost conformal there.

The ratio, $\zeta/\eta$ decreases as we increase the temperature. The decrease is quite steeper until 
we reach $T=2.0 T_c$. For higher values of $T$ it is much slower. It is hard to make clear
 cut statement in regard to the behavior of $\zeta/\eta$ with temperature, since by looking at 
Eqs. (23) and (24), it is clear that the behavior of $\zeta/\eta$ as a function of temperature is 
mainly governed by the temperature dependence of trace anomaly (through quasi-gluon dispersion relation),
speed of sound, $c_s^2$ and temperature dependence of $z_g$ and gluon quenching parameter, $\hat{q}$. 

To compare the perturbative QCD prediction of the ratio $\zeta/\eta$,
we consider $R_{pert}\equiv \frac{\zeta}{\eta (c_s^2-\frac{1}{3})^2}$, where $(c_s^2-\frac{1}{3})$ 
can be thought of as the measure of conformal symmetry, which we call conformal measure.  
For scalar field theories, $\zeta/\eta= 15 (c_s^2-\frac{1}{3})^2$~\cite{scalar}, and this has been found 
to be true for a photon gas coupled with hot matter by Weinberg~\cite{weinberg}. The pre-factor $15$ is not
fixed for perturbative QCD but the scaling $\zeta/\eta\sim (c_s^2-\frac{1}{3})^2$is valid\cite{arnold}). Note that 
in certain strongly coupled near conformal theories with gravity dual the ratio $\zeta/\eta$ shows linear dependence on the conformal measure~\cite{confo}. To compare with the latter, we consider the ratio 
$R_{str}\equiv\frac{\zeta}{\eta (c_s^2-\frac{1}{3})}$. We have shown the behavior of $R_{pert}$ and $R_{str}$ as a function of temperature in Fig. 2. Clearly, none of these two scaling are respected by the ratio $\zeta/\eta$ in Fig. 2
even at $2.5\ T_c$. It is safer to say that $\zeta/\eta$ for LEOS which is obtained from transport equation with 
Vlasov-Dupree term~\cite{bmuller,chandra_eta2} neither shows linear nor the quadratic dependence with the conformal measure, $(c_s^2-\frac{1}{3})$. However, one can realize the quadratic scaling of $\zeta/\eta$ with the conformal measure 
in a certain limiting case. It is easy to say from Eqs.(24) and (25) that for  $E_p=p$ ($p<<T^2\partial_T (ln(z_g))$), if 
the thermal distribution of quasi-gluons shows near ideal behavior, and with constant value of $\hat{q}/T^3$, the 
quadratic scaling can be achieved. Moreover, this may perhaps be realized at higher temperatures which are not relevant for QGP in RHIC and LHC. If we compare qualitatively 
our prediction of $R_{pert}$, and $R_{str}$ with the leading order result of~\cite{chen_bulk} (see Fig. 4 of this Ref.) , we find opposite 
trend of these quantities at very high temperature. The former decreases, although slowly, in contrast to the
latter, as a function of temperature. This could perhaps be because of their origin from the distinct physical processes in hot QCD medium. The slow decreases of the former at higher temperatures, could be understood as the effect of thermal distribution function of quasi-gluons (through $z_g$, since $z_g$ increases very slowly as a function of temperature, and will asymptotically approach to unity).

Finally, in Fig. 3, we have shown the temperature dependence of $\zeta/s$ and $\eta/s$. The $\zeta/s$ decreases as 
with increasing temperature for $T\geq 1.5$, in contrast to $\eta/s$. As mentioned earlier, $\zeta/s$ and $\eta/s$
becomes equal around $1.5 T_c$ (below which $\zeta/s$ is higher, and lower for higher temperatures.). Again the behavior 
is predominantly controlled by the behavior of $c_s^2$, and the  trace anomaly through the modified dispersion relation with temperature.

\section{Conclusions and future prospects}
In conclusion, we have estimated the temperature dependence of 
bulk viscosity to entropy density ratio ($\zeta/s$), and bulk viscosity relative to 
shear viscosity, $\zeta/\eta$  within a quasi-particle model for pure glue   QCD at high temperature 
by employing transport theory. We have  determined $\zeta/\eta$, exactly and unambiguously. In our analysis, 
these quantities get contributions from the instabilities in the chromo-electromagnetic fields due
 to the anisotropic thermal distribution of the partons in QGP. The mechanism has succeeded in explaining 
the small $\eta/s$ and large value of the ratio $\zeta/\eta$. In fact, $\zeta/\eta$ is around $2.3$ at $1.3\ T_c$,
of the order of unity at $1.5\ T_c$, and 0.2 at $2\ T_c$. This tells us that the breaking of conformal symmetry
 in hot QCD plays crucial role even at $2\ T_c$. In consequence, shear and bulk viscosities are equally
 important while studying the hydrodynamic evolution of QGP at RHIC and LHC.
One cannot simply ignore bulk viscosity even at $2.0 T_c$ while modeling the heavy ion collisions.
Moreover, $\eta/s$ increases as a function of temperature, in contrast to $\zeta/s$ beyond $1.5 T_c$. 
As expected $\zeta/s$ and $\zeta/\eta$ are vanishingly small beyond $2.5 T_c$. This may be due to the 
fact that conformal measure is very small there, and the speed of sound is closer to $1/3$.
We have compared our predictions on $\zeta/\eta$ to the leading order result on the same quantity
obtained by~\cite{chen_bulk}. Interestingly, in the perturbative region (temperatures beyond $1.5 T_c$), our study also agree with the near conformal picture of hot QCD similar to leading order results of Chen {\it et. al}~\cite{chen_bulk}.
On the other hand the predictions are in contrast at lower temperatures. However, this may not be 
thought of as the complete story, an adequate analysis on the interplay of our predictions on $\zeta$, and leading order 
prediction is very much desired, and will be a matter of future investigations.   

We have addressed the temperature dependence of the bulk and shear viscosities of pure glue sector of hot QCD only. 
An extension to full QCD including collision term, employing the understanding of~\cite{vinod}, will be a matter of
future investigations. We strongly believe that a similar analysis will also be valid in the case of full QCD. 
The most interesting study would be to include the temperature dependence of $\eta/s$ and $\zeta/s$
in the existing hydro codes to model QGP, and see how various observables get modifications.
Moreover, future directions may include exploration on the effects of $\eta$ and $\zeta$ 
on the quarkonia suppression in heavy ion collisions along the lines of~\cite{chandra_iitr,dumitru}. 
Finally, it would be of interest to include the baryon chemical potential utilizing 
the very recent lattice studies~\cite{karsch,petre}, and determine the transport coefficients.

\vspace{3mm}
\noindent{\bf Acknowledgements:}
VC is thankful to Prof. F. Karsch, and Prof. Saumen Datta for providing the lattice QCD data, Prof. Rajeev Bhalerao, Prof. V. Ravishankar  for invaluable suggestions and encouragement, and Dr. Sudhansu  Biswal for the numerical help.
VC would like to thank Prof. U. A. Wiedemann for invaluable discussions at the CERN-Theory division, and sincerely acknowledge the hospitality of CERN-Theory Division, CERN, Geneva through the CERN visitor program.


\begin{thebibliography}{99}
\bibitem{expt} STAR collaboration, J. Adams {\it et al}., Nucl. Phys. {\bf A 757},
102 (2005); PHENIX Collaboration, Nucl. Phys. {\bf A 757}, 184 (2005);
 PHOBOS Collaboration, Nucl. Phys. {\bf A 757}, 28 (2005);
 BRAHMS Collaboration , Nucl. Phys. {\bf A 757}, 1 (2005).

\bibitem{flow_rhic} STAR collaboration, J. Adams {\it et al}., Nucl. Phys. {\bf A 757}, 102 (2005).

\bibitem{flow_lhc}
Derek Teaney, Li Yan, {\tt arXiv:1010.1876 [nucl-th]}; Burak Han Alver, Clement Gombeaud, Matthew Luzum, Jean-Yves Ollitrault, Phys. Rev. {\bf C 82}, 034913 (2010). 

\bibitem{bhalerao}
Rajeev S. Bhalerao, Matthew Luzum, Jean-Yves Ollitrault, {\tt arXiv:1106.4940[nucl-ex]};
{\tt arXiv:1104.4740[nucl-th]}; Matthew Luzum, Jean-Yves Ollitrault Phys. Rev. Lett. {\bf 106}, 102301 (2011). 

\bibitem{alice}
M. Krzewicki, for the ALICE Collaboration, QM-2011, {\tt arXiv:1107.0080v1 [nucl-ex]}.

\bibitem{lhc} K. Aamodt {\it et. al}[The Alice Collaboration],
{\tt arXiv:1011.3914 [nucl-ex]};  Phys. Rev. Lett. {\bf 105}, 252301 (2010); Phys. Rev. Lett. {\bf 106}, 032301
(2011).

\bibitem{shrvis} H. B. Meyer, Phys. Rev. {\bf D 76}, 10171 (2007); Lacey {\it et. al}, Phys. Rev. Lett. {\bf 98}, 092301 (2007); Zhe Xu and Carsten Greiner, Phys. Rev. Lett. {\bf 100},172301 (2008); Zhe Xu, Carsten Greiner, Horst Stoecker, Phys. Rev. Lett. {\bf 101}, 082302 (2008); Adare {\it et. al}, Phys. Rev. Lett. {\bf 98}, 172301 (2007); Sean Gavin and Mohamed Abdel-Aziz, Phys. Rev. Lett. {\bf 97}, 162302 (2006); Alex Buchel, Phys. Lett. {\bf B 663}, 286 (2008);
P. Kovtun, D.T. Son, A. O. Starinets, Phys. Rev. Lett.
{\bf 94}, 111601 (2005).

\bibitem{khz1} D. Kharzeev, K. Tuchin, JHEP {\bf 0809}, 093 (2008).

\bibitem{khz2} F. Karsch, D. Kharzeev, K. Tuchin,
 Phys. Lett. {\bf B 663}, 217 (2008). 

\bibitem{chandra_eta1} Vinod Chandra and V. Ravishankar, Euro. Phys. J C {\bf 59}, 705 (2009).

\bibitem{chandra_eta2} Vinod Chandra and V. Ravishankar, Euro. Phys. J C {\bf 64}, 63 (2009).


\bibitem{moore}
Guy D. Moore, Omid Saremi, JHEP {\bf 0809}, 015 (2008).
\bibitem{meyer} H. B. Meyer, Phys. Rev. Lett. {\bf 100}, 162001 (2008). 
\bibitem{tmr} D. Teaney, Phys. Rev. {\bf D 74}, 0450125 (2006) ({\tt hep-ph/0602044});
Guy D. Moore, Omid Saremi, JHEP {\bf 0809}, 015 (2008) ({\tt arXiv:0805.4201[hep-ph]});
P. Romatschke, D. T. Son, Phys. Rev. {\bf D 80}, 065021 (2009) ({\tt arXiv:0903.3946}).
\bibitem{meyer1} H. B. Mayer, JHEP {\bf 1004}, 099 (2010) ({\tt arXiv:1002.3343[hep-lat]}).

\bibitem{nakamura}
Atsushi Nakamura, Sunao Sakai, Phys. Rev. Lett. {\bf 94}, 072305 (2005);
Nucl. Phys. {\bf A 774}, 775 (2006). 

\bibitem{heinz} Huichao Song, Ulrich W Heinz,  Phys. Rev. C {\bf 81}, 024905 (2010).

\bibitem{den} G. S. Denicol, T. Kodama, T. Koide, Ph. Mota, 
     Phys. Rev. {\bf C 80}, 064901 (2009).

\bibitem{raj1} 
    G. S. Denicol, T. Kodama, T. Koide, Ph. Mota, Nucl. Phys. {\bf A 830},729c (2009).

\bibitem{hirano}   
 Akihiko Monnai, Tetsufumi Hirano, Nucl. Phys. {\bf A 830}, 471c (2009;  Phys. Rev. {\bf C 80}, 054906 (2009).

\bibitem{raj}
Krishna Rajagopal, Nilesh Tripuraneni, JHEP {\bf 1003}, 018 (2010);
Jitesh R. Bhatt, H. Mishra,  V. Sreekanth {\tt arXiv:1103.4333}.

\bibitem{efaaf} 
M. J. Efaaf, Zhong-Qian Su, Wei-Ning Zhang, {\tt arXiv:1008.1531}.

\bibitem{pion}
D.Fernandez-Fraile, A.Gomez Nicola, Phys. Rev. Lett. {\bf 102}, 121601 (2009).

\bibitem{fries}   
 Rainer J. Fries, Berndt M\"uller, Andreas Sch\"afer,
 Phys. Rev. {\bf C 78}, 034913 (2008).

\bibitem{torri}   Giorgio Torrieri, Boris Tomasik, Igor Mishustin
Phys. Rev. {\bf C 77}, 034903 (2008); Acta. Phys. Polon. {\bf B 39}, 1733 (2008). 

\bibitem{boz} Piotr Bozek, Phys. Rev. {\bf C 81}, 034909 (2010).

\bibitem{cosmo} A. Tawfik , M. Wahba,  H. Mansour, T. Harko,
{\tt arXiv:1008.0971}; Arturo Avelino, Ulises Nucamendi, JCAP {\bf 1008}, 006  (2010).

\bibitem{sm} Xinyang Wang, Igor A. Shovkovy, {\tt arXiv:1006.1293}; Shou-wan Chen, Hui Dong, Qun Wang,  J. Phys. G: Nucl. Part. Phys. {\bf 36}, 064050 (2009). 

\bibitem{ns} Brynmor Haskell, Nils Andersson, 	{\tt arXiv:1003.5849}; Massimo Mannarelli, Cristina Manuel, Phys. Rev. {\bf D 81}, 043002 (2010); Xu-Guang Huang, Mei Huang, Dirk H. Rischke, Armen Sedrakian,  Phys. Rev. {\bf D 81}, 045015 (2010).

\bibitem{shhydro} 
Matthew Luzum, Paul Romatschke, Phys. Rev. {\bf C 78}, 034915 (2008).

\bibitem{bulkhydro}
 Huichao Song, Ulrich W. Heinz, Nucl. Phys. {\bf A 830}, 467c (2009).

\bibitem{sakai}
   C. Sasaki, K. Redlich,  Phys. Rev. {\bf C 79}, 055207 (2009).

\bibitem{quasi1} P. Chakraborty, J. I. Kapusta, {\tt arXiv:1006.0257}.
  
\bibitem{bmuller} Masayuki Asakawa, Steffen A. Bass and Berndt M\"{u}ller, Prog. Theor. Phys. {\bf 116}, 725
(2007).

\bibitem{bmuller1} Masayuki Asakawa, Steffen A. Bass and Berndt M\"{u}ller, Phys. Rev. Lett. {\bf 96}, 252301 (2006);
 Abhijit Majumdar, Berndt M\"{u}ller and Xin-Nian Wang, Phys. Rev. Lett. {\bf 99}, 192301 (2007).

\bibitem{chromw}
 S. Mrowczynski,  Phys. Rev. {\bf C 49}, 2191 (1994); M Strickland, Braz. J. Phys. {\bf 37}, 762 (2007); {\tt hep-ph/0611349}; P. Arnold and G. Moore, Phys. Rev. {\bf D 73}, 025013 (2006).

\bibitem{pratt}
Kerstin Paech, Scott Pratt, Phys. Rev. {\bf C 74}, 014901 (2006).

\bibitem{dupree}
T. H. Dupree, Phys. Fluids {\bf 9}, 1773 (1966); {\it ibid.} {\bf 11} 2680 (1968).

\bibitem{niu} T. Abe, K. Niu,  J. Phys. Soc. Japan {\bf 49} 717 (1980); {\it ibid.} {\bf 49} 725, (1980).

\bibitem{bmuller2}
M. Asakawa, S. A. Bass, B. M\"uller, J. Phys. G {\bf 34}, S839 (2007).

\bibitem{weibel}
E.S. Weibel, Phys. Rev. Lett. {\bf 2}, 83 (1959).

\bibitem{sma} S. M\`rowcz\`ynski, Phys. Lett. {\bf B 214}, 587 (1988); {\it ibid.} {\bf B 314}, 118 (1993);
 P. Romatschke, M. Strickland, Phys. Rev. {\bf D 68} 036004, (2003).

\bibitem{kar} Our sincere thanks to F. Karsch for providing us the lattice data for pure $SU(3)$ gauge theory in the past which has been quoted in~\cite{chandra_eta2}, due to which the present analysis became possible.

\bibitem{lat_eos}
G. Boyd {\it et. al}, Phys. Rev. Lett. {\bf 75}, 4169 (1995); Nucl.
Phys. {\bf B 469}, 419 (1996); M. Panero, Phys. Rev. Lett. {\bf 103}, 232001 (2009); 
F. Karsch, E. Laermann, A. Peikert, Phys. Lett. {\bf B 478},
447 (2000); M. Cheng {\it et. al}, Phys. Rev. {\bf D 77}, 014511 (2008); A. Bazavov {\it et. al}, Phys. Rev. {\bf D 80}, 014504 (2009); M. Cheng {\it et. al}, Phys. Rev. {\bf D 81},054504 (2010).

\bibitem{lat_eos1} Szabolcs Borsanyi {\it et. al}, JHEP {\bf 1009},073 (2010); JHEP {\bf 11}, 077 (2010); Y. Aoki
{\it et al.}, JHEP {\bf 0601}, 089 (2006); JHEP {\bf 0906}, 088 (2009).

\bibitem{vinod} Vinod Chandra, V. Ravishankar, Phys. Rev. {\bf D 84}, 074013 (2011) ( {\tt arXiv:1103.0091 [nucl-th]}).

\bibitem{vinod_quasi} Vinod Chandra , R. Kumar, V. Ravishankar, Phys.Rev. {\bf C 76}, 054909 (2007); Indian J. Phys. {\bf 84}, 1789 (2010); Vinod Chandra, A. Ranjan, V. Ravishankar, Eur. Phys. J. {\bf A 40}, 109 (2009); {\tt arXiv:0801.1286[hep-ph]}.  

\bibitem{vinod_quasi1} Vinod Chandra, V. Ravishankar, Nucl. Phys. {\bf A 848}, 330 (2010).

\bibitem{kamp} A. Peshier, B. K\"{a}mpfer, G. Soff, Phys. Rev. {\bf C 61}, 045203
(2000); Phys. Rev. {\bf D 66}, 094003 (2002).

\bibitem{pesh} A. Peshier {\it et. al}, Phys. Lett. {\bf B 337}, 235 (1994); Phys. Rev. {\bf D 54}, 2399 (1996).

\bibitem{bannur} Vishnu M. Bannur, Phys. Rev. {\bf C 75}, 044905 (2007); {\it ibid.} {\bf C 78}, 045206 (2008); 
 JHEP {\bf 0709}, 046 (2007).

\bibitem{rebhan} A. Rebhan, P. Romatschke, Phys. Rev. {\bf D 68}, 0250022 (2003).

\bibitem{thaler}  M. A. Thaler, R. A. Scheider, W. Weise, Phys. Rev. {\bf C 69}, 035210 (2004);
 K. K. Szab\`{o}, Anna I. T\`{o}th, JHEP {\bf 06}, 008 (2003).


\bibitem{polyakov} A. Dumitru and R. D. Pisarski, Phys. Lett. {\bf B 525}, 95
(2002); K. Fukushima, Phys. Lett. {\bf B 591}, 277 (2004); S. K. Ghosh {\it et. al}, Phys. Rev. {\bf D 73}, 114007 (2006);
 H. Abuki, K. Fukushima, Phys. Lett. {\bf B 676}, 57 (2006); H. M. Tsai, B Muller, J. Phys. {\bf G 36}, 075101 (2009).

\bibitem{ella} M. D'Elia, A. Di Giacomo, E. Meggiolaro, Phys. Lett. {\bf B 408}, 315 (1997); Phys. Rev. {\bf D 67}, 114504 (2003).

\bibitem{casto} P. Castorina, M. Mannarelli, Phys. Rev. {\bf C75}, 054901 (2007); 
Phys. Lett. {\bf B 664}, 336 (2007).

\bibitem{bluhm}
M. Bluhm, B.K\"{a}mpfer, K. Redlich, {\tt arXiv:1011.5634(nucl-th)}; {\tt arXiv:1101.3072[nucl-th]}.

\bibitem {aks} A. S. khvorostukhin, V. D. Toneev, D. N. Voskresersky, 
Phys. Rev. {\bf C 83}, 035204 (2011); Santosh K. Das, Jan-e Alam,
Phys. Rev. {\bf D 83}, 114011 (2011). 

\bibitem{landau} E. M. Lifshitz and L. P. Pitaevskii, Physical Kinetics(Landau
and Lifshitz; Volume 10) Pergamon Press.

\bibitem{gupta}
Rajiv V. Gavai, Sourendu Gupta, Swagato Mukherjee, PoS LAT {\bf 2005}, 173 (2005).

\bibitem{zantow} O. Kaczmarek, F. Karsch, P. Petreczky, F. Zantow,
Phys. Lett. {\bf B 543}, 41 (2002); Phys. Rev. {\bf D 70}, 074505
(2004); Olaf Kaczmarek, Felix Zantow, Phys. Rev. {\bf D 71},
114510 (2005).

\bibitem{chen_bulk} Jiunn-Wei Chen, Jian Deng, Hui Dong, Qun Wang, 
{\tt arXiv:1107.0522v2[hep-ph]}.

\bibitem{chen_shear}
Jiunn-Wei Chen, Jian Deng, Hui Dong, Qun Wang,
Phys. Rev. {\bf D 83}, 034031 (2011); {\it ibid.} {\bf D 84}, 0399902(E) (2011).

\bibitem{hatq} Hanzhong Zhang, J. F. Owens, Enke Wang, Xin-Nian
Wang, Phys. Rev. Lett. {\bf 98}, 212301 (2007); A. Majumder,
C. Nonaka, S. A. Bass, Phys. Rev. {\bf C 76}, 041902 (2007); Peign, D.Schiff,
Nucl. Phys. {\bf B 483}, 291 (1997); N. Armesto, L. Cunqueiro,
C. A. Salgado, W.-C. Xiang, JHEP {\bf 0802}, 048 (2008).

\bibitem{arnold} P. Arnold, C. Dolan, Guy D. Moore, Phys. Rev. {\bf D 74}, 085021 (2006).

\bibitem{confo} P. Benincasa, A. Buchal, A. O. Strarinets, Nucl. Phys. {\bf B 733}, 160 (2006);
A. Buchal, Phys. Rev. {\bf D 72}, 106002 (2005).

\bibitem{scalar} R. Horsley, W. Schoenmaker, Nucl. Phy. {\bf B 280}, 716 (1987).

\bibitem{weinberg} S. Weinberg, Astrophys. {\bf J 168}, 175 (1971). 

\bibitem{chandra_iitr}
V. Agotiya, Vinod Chandra, B. K. Patra, Phys. Rev. {\bf C 80}, 025210 (2009); Euro. Phys. J {\bf C 67}, 465 (2010).


\bibitem{dumitru} Adrian Dumitru, Yun Guo, Agnes Mocsy, M. Strickland, Phys. Rev. {\bf D 79} 054019 (2009).

\bibitem{karsch} Frithjof Karsch, Bernd-Jochen Schaefer, Mathias Wagner, Jochen Wambach, Phys. Lett. {\bf B 698}, 256 (2011). 

\bibitem{petre} Pasi Huovinen, Peter Petreczky, QM-2011, {\tt arXiv:1106.6227 [nucl-th]}.

\end{thebibliography}
\end{document}